\documentclass[12pt]{article}
\usepackage{amsmath}
\usepackage{amsfonts}
\usepackage{amssymb}
\usepackage{mathrsfs}
\usepackage{multicol}
\usepackage{color}
\usepackage{graphicx}
\usepackage{pstricks}
\usepackage{pst-node}
\usepackage{wrapfig}
\usepackage{enumerate}
\usepackage{caption}
\usepackage{cite}
\usepackage{subfigure}
\usepackage{amsfonts,amssymb}
\usepackage{amsthm}
\usepackage{amscd}
\definecolor{gray1}{gray}{0.1}
\definecolor{gray2}{gray}{0.2}
\definecolor{gray3}{gray}{0.3}
\definecolor{gray4}{gray}{0.4}
\definecolor{gray5}{gray}{0.5}
\definecolor{gray6}{gray}{0.6}
\definecolor{gray7}{gray}{0.7}
\definecolor{gray8}{gray}{0.8}
\definecolor{gray9}{gray}{0.9}

\newpsobject{showgrid}{psgrid}{subgriddiv=1,griddots=10,gridlabels=6pt}

\definecolor{dark-green}{rgb}{0,0.7,0}
\definecolor{dark-blue}{rgb}{0,0.2,0.5}
\definecolor{med-blue}{rgb}{0,0.7,1}
\definecolor{mblue}{rgb}{0,0.2,1}
\definecolor{cnc}{rgb}{0.8,0,0}
\definecolor{light-red}{rgb}{1,0.8,0.8}
\definecolor{dark-yellow}{rgb}{1,0.8,0}
\definecolor{light-blue}{rgb}{0.8,0.9,1}
\definecolor{verylight-blue}{rgb}{0.93,0.95,1}
\definecolor{light-yellow}{rgb}{1,0.9,0.8}
\definecolor{grey}{gray}{0.88}

\newpsobject{showgrid}{psgrid}{subgriddiv=1,griddots=10,gridlabels=6pt}

\begin{document}

\thispagestyle{empty}

\setlength{\abovecaptionskip}{10pt}

\begin{center}
{\Large\bfseries\sffamily{Charged scalar-tensor solitons and black holes \\
with (approximate) Anti-de Sitter asymptotics}}
\end{center}
\vskip 3mm
\begin{center}
{\bfseries{\sffamily{Yves Brihaye$^{\rm 1}$ and Betti Hartmann$^{\rm 2}$}}}\\
\vskip 3mm{$^{\rm 1}$\normalsize Physique-Math\'ematique, Universit\'e de Mons-Hainaut, 7000 Mons, Belgium}\\
{$^{\rm 2}$\normalsize{Instituto de F\'isica de S\~ao Carlos (IFSC), Universidade de S\~ao Paulo (USP), CP 369, \\
13560-970 , S\~ao Carlos, SP, Brazil}}
\end{center}

\begin{abstract} 
We discuss charged and static solutions in a
shift-symmetric scalar-tensor gravity model including a negative cosmological constant. The solutions
are only approximately Anti-de Sitter (AdS) asymptotically.  While spherically symmetric black holes with scalar-tensor hair do exist in our model, the uncharged spherically symmetric scalar-tensor solitons constructed recently cannot be generalised to include charge. We point out that this is due to
the divergence of the electric monopole at the origin of the coordinate system, while higher order multipoles are well-behaved. 
We also demonstrate that black holes with scalar hair exist only for horizon value 
larger than that of the corresponding {\it extremal} Reissner-Nordstr\"om-AdS (RNAdS) solution, i.e. that we cannot construct solutions with arbitrarily small horizon radius.
We demonstrate that for fixed $Q$ a horizon radius exists at which the  specific heat $C_Q$ diverges - signalling a transition from thermodynamically unstable to stable black holes. In contrast to the RNAdS case, however, we have only been able to construct a stable phase of large horizon black holes, while a stable phase of small horizon black holes does not (seem to)  exist.
\end{abstract}
 
\section{Introduction}
Already included in the vacuum solution to the Einstein equation presented by Schwarzschild in 1916 \cite{schwarzschild}, serious
research on objects so dense that a light signal emitted from their surface, the so-called {\it event horizon}, becomes infinitely redshifted for an external observer, began only in the late 1960. Later coined {\it black holes} these objects were predicted to appear in the gravitational collapse of
very heavy stars \cite{penrose}.  Very early on in the studies of the available analytical solutions, it had been realized that asymptotically flat black hole solutions of the (electro-)vacuum
Einstein(-Maxwell) equations appear to be very structureless objects determined uniquely by a small
number of parameters only~: their mass (, charge) and angular momentum. A number of uniqueness
theorems were proven in the following \cite{BHuniqueness} and it had been conjectured that {\it black hole
have no hair} also in more general situations. However, this conjecture has been falsified by the
construction of a number of counterexamples, such as e.g. black holes in models with non-linear matter sources \cite{hairyBHs} and
black holes in asymptotically Anti-de Sitter (AdS) space-time \cite{BHAdS}.
Another intriguing property of black holes is the analogy of the laws describing their behaviour to that of the laws of Thermodynamics \cite{hawking_bardeen_carter}. Being a pure analogy in classical General Relativity, the surface gravity and area of the black hole horizon correspond to temperature and entropy. Using a semiclassical approach it was shown that this is, indeed, more than a mere analogy and that black holes do emit thermal radiation from their horizon \cite{hawking1974}, which, however, for stellar mass black holes would be undetectably small and orders of magnitude smaller than that of the Cosmic Microwave background (CMB) radiation.
Recent observations of collisions of two black holes with the help of gravitational waves \cite{Abbott:2016blz,
Abbott:2016nmj,Abbott:2017vtc,Abbott:2017oio,Abbott:2017gyy} demonstrated that not only does
General Relativity remain the best theory we have for the gravitational interaction to this day, but also that astrophysical black holes seem to be
simple objects, indeed. For once, the  collisions only produced gravitational
and no other radiation, while the recently detected collision of two neutron stars has also been observed in terms of a matching gamma-ray burst \cite{Monitor:2017mdv,TheLIGOScientific:2017qsa,Troja:2017nqp}.
Moreover, the event and final state of the collision are extremely well matched by templates generated with black holes described only by their mass and angular momentum. The question then remains why one should be interested to study
extended gravity models and black hole solutions which do not fulfil the No-hair conjecture. The answer is two-fold: (a) in order to exclude certain extended gravity models using the future observational data available, we need to make predictions about the properties of black hole (and other type of) solutions in these models and (b) classical General Relativity (and its extensions) in asymptotically AdS space-time is suggested to be dual to a strongly coupled (conformal) quantum field theory (CFT) without gravity  on the boundary of AdS within the
context of the AdS/CFT correspondence, or more generally within the gauge-gravity duality \cite{adscft,ggdual}. This opens the possibility to find models for phenomena that are -- up to now -- very difficult to access from the QFT side. Examples are high-temperature superconductivity \cite{hhh,reviews} as well as certain aspects of Quantum Chromodynamics \cite{Ammon:2015wua}. In particular, the gauge-gravity duality
allows the identification of the surface gravity of the classical black hole solution with the temperature of the dual field theory (up to a factor). 
Interestingly, black holes in space-time including a cosmological constant $\Lambda$ have also been used in {\it black hole chemistry} to describe e.g. phase transitions in Van-der-Waals gases \cite{Chamblin:1999tk, Chamblin:1999hg, Wu:2000id, Kubiznak:2014zwa,Mann:2016trh,Kubiznak:2016qmn}. The idea is to
generalize the 1st law of black hole mechanics -- the analogue of the 1st law of Thermodynamics --
to include a pressure work term of the form $V dp$  by identifying the cosmological constant $\Lambda$ with the pressure $p$ via $p=-\Lambda/(8\pi)$ and
the volume $V$ with the thermodynamic volume (for a discussion on the interpretation of the latter see e.g. \cite{Kubiznak:2016qmn}). 

Scalar-tensor gravity models have gained lots of interest in recent years (for reviews see e.g. \cite{Deffayet:2013lga, Charmousis:2014mia}), in particular in the context of so-called
Horndeski theories \cite{horndeski} which -- like General Relativity -- lead to field equations that
are maximally second order in derivatives. Asymptotically flat black holes have been discussed
extensively \cite{Babichev:2017ab} and a explicit example of a black hole with scalar-tensor hair has been presented \cite{Sotiriou:2014pfa}. Unfortunately, a big part of these models are now ruled out by the recent gravitational wave observations \cite{Ezquiaga:2017ekz}. 
However, this does not exclude the use of these gravity models in the context of the gauge-gravity duality.
This is the approach we are taking in this paper. We discuss charged and static
black hole and soliton solutions of a scalar-tensor gravity model in (3+1) space-time dimensions including a
negative cosmological constant. 
Our paper is organised as follows~: in Section 2, we give the model and Ansatz. In Section 3, we discuss
soliton solutions and demonstrate that static, spherically symmetric solutions do not exist. In Section 4, we 
discuss charged black holes with scalar hair and point out that black holes can only be constructed for a sufficiently large value of the horizon radius. Section 5 contains our conclusions.

\section{The model}
The model we are studying in this paper is a Horndeski scalar-tensor model coupled to a $U(1)$ gauge field.  Its action
reads~: 
\begin{equation}
\label{action}
S =  \int  {\rm d}^4x  \sqrt{-g} \left[R-2\Lambda  +  \frac{\gamma}{2}  \phi {\cal G} -  
\partial_{\mu} \phi  \partial^{\mu} \phi   - \frac{1}{4} F_{\mu\nu} F^{\mu\nu} \right]   \ ,
\end{equation}
where the Gauss-Bonnet term ${\cal G}$ is given by
\begin{equation}
 {\cal G} = R^{\mu \nu \rho \sigma} R_{\mu \nu \rho \sigma} - 4 R^{\mu \nu}R_{\mu \nu} + R^2  \ ,
\end{equation}
and $F_{\mu\nu}=\partial_{\mu} A_{\nu} - \partial_{\nu} A_{\mu}$ is the field strength tensor of the $U(1)$ gauge field $A_{\mu}$. $\gamma$ is the scalar-tensor coupling and $\Lambda < 0$ is the negative valued cosmological constant. Units are chosen such that $16\pi G\equiv 1$.
 
Varying the  action (\ref{action}) with respect to the metric, the scalar field and the gauge field gives the following coupled system of non-linear differential equations~:
\begin{equation}
\label{eom1}
G_{\mu\nu} +\Lambda g_{\mu\nu} - \partial_{\mu} \phi \partial_{\nu} \phi + \frac{1}{2}g_{\mu\nu} \partial_{\alpha} \phi \partial^{\alpha} \phi + \frac{\gamma}{2}( g_{\rho\mu} g_{\sigma\nu} + g_{\rho\nu} g_{\sigma\mu})\nabla_{\lambda} (\partial_{\gamma} \phi \epsilon^{\gamma \sigma\alpha\beta} \epsilon^{\delta\eta } R_{\delta\eta\alpha\beta})  = 0 \ ,
\end{equation}
\begin{equation}
\label{eom2} 
\square \phi = - \frac{\gamma}{2} {\cal G} \  , \ \ \partial_{\mu} \left(\sqrt{-g} F^{\mu\nu}\right) = 0 \ .
\end{equation}
For the metric and the scalar, we choose the following Ansatz~:
\begin{equation}
\label{ansatz}
ds^2=-N(r) \sigma(r)^2 dt^2 + \frac{1}{N(r)} dr^2  + r^2 \left(d\theta^2 + \sin^2 \theta d \varphi^2\right)  \ \ , \ \  \phi=\phi(r)  \  \ .
\end{equation}
We will discuss the Ansatz for the U(1) gauge field separtely in the case of
solitons and black holes. In \cite{Brihaye:2017wln} the case with vanishing gauge field, i.e. the uncharged case has
been studied.  It was found that the metric and scalar functions asymptotically behave as follows~:
\begin{equation}
\label{expansion_infinity}
N(r) = C_1 - \frac{\lambda}{3} r^2 - \frac{M}{r} + {\rm O}(r^{-2}) \ \ , \ \ \sigma(r)=1 + {\rm O}(r^{ -2}) \ \ , \ \
\phi(r) = C_0 - C_2 \ln(r) + {\rm O}(r^{-2})   \ , 
\end{equation}
where $M$ determines the gravitational mass  of the solution.
${\lambda}$,  $C_0$. $C_1$, $C_2$ are constants that depend on $\gamma$ and $\Lambda$.  Note that the
non-minimal scalar-tensor coupling leads to a renormalisation of the cosmological constant and that the
leading order term is of Anti-de Sitter form. However, the constant $C_1\neq 1$ in general for $\gamma\neq 0$
and hence the metric does not approach global Anti-de Sitter (with a rescaled cosmological constant) asymptotically.
The metric is hence only {\it approximately Anti-de Sitter}.

\section{Soliton solutions}
We have investigated the coupling of the static, spherically symmetric scalar-tensor
solitons to a spherically symmetric, static electric field, but were unable to find
a corresponding globally regular solution numerically.

Hence, we have studied the electric and magnetic multipoles of a static and axially symmetric electromagnetic field in the background
of the scalar-tensor soliton solution presented in \cite{Brihaye:2017wln}. We will show in the following that the
spherically symmetric monopole always diverges in this case, hence spherically symmetric, charged generalisations of the AdS scalar-tensor solitons constructed in \cite{Brihaye:2017wln} do not exist.
 
Following \cite{Herdeiro:2015vaa} we choose the static, axially symmetric separation Ansatz for the U(1) gauge field as follows
\begin{equation}
A_{\mu} dx^{\mu} = A_t(r,\theta) dt + A_{\varphi}(r,\theta) d\varphi  \ .
\end{equation}
The Maxwell equations (see (\ref{eom2}))  then read~:
\begin{equation}
N \sigma \partial_r \left(\frac{r^2 \partial_r A_t}{\sigma}\right) + \frac{1}{\sin\theta} \partial_{\theta} \left(\sin\theta \partial_{\theta} A_t\right) = 0    \ \  \  , \ \ \  
\frac{r^2}{\sigma} \partial_r \left(r^2 N \sigma \partial_r A_{\varphi}\right) + \sin\theta \partial_{\theta} \left(
\frac{\partial_{\theta} A_{\varphi}}{\sin\theta}\right) = 0     \  .
\end{equation}
These equations have been studied in great detail for $\gamma=0$, i.e. in global AdS space-time in \cite{Herdeiro:2015vaa}, where $\sigma(r)\equiv 1$ and $N(r)=1- \frac{\Lambda}{3}r^2$. In this latter case, analytical solutions are possible, which is not the case when extending the
model to include a non-minimally coupled scalar field. 

In the following, we will treat the electric and magnetic multipoles separately. Of course, solutions with both an electric
and a magnetic field are possible, as was shown for the case of global AdS in \cite{Herdeiro:2015vaa}.
\clearpage
\subsection{Electric multipoles}

Choosing $A_{\varphi}\equiv 0$, we employ the following separation of variables
\begin{equation}
A_{t}(r,\theta) = R_{\ell}(r) {\cal P}_{\ell}(\cos\theta) 
\end{equation}
where the ${\cal P}_{\ell}(\cos\theta)$, $\ell=0,1,2,3...$ are the Legendre polynomials.  Introducing the dimensionless quantities
\begin{equation}
x=\sqrt{-\frac{\Lambda}{3}} r \  \ \ \ , \ \ \ \ \zeta^2=\frac{\gamma^2\Lambda^2}{9} 
\end{equation}
the equation for $R_{\ell}$ reads (prime denotes the derivative with respect to $x$)~:
\begin{equation}
\label{eq1}
R_{\ell}''=\ell(\ell+1)\frac{R_{\ell}}{x^2 N} + R_{\ell}' \left(\frac{\sigma'}{\sigma} - \frac{2}{x}\right)
\end{equation}
with the metric functions $N$ and $\sigma$ given by the numerical solutions presented in \cite{Brihaye:2017wln}.

\begin{figure}[ht!]
\begin{center}
\input{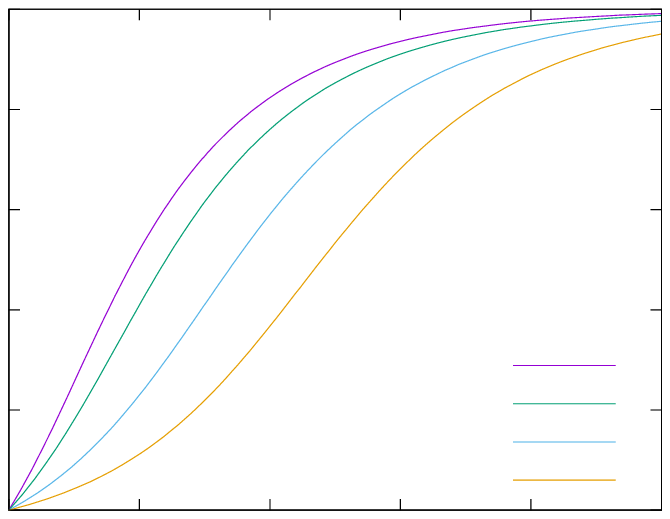}
\input{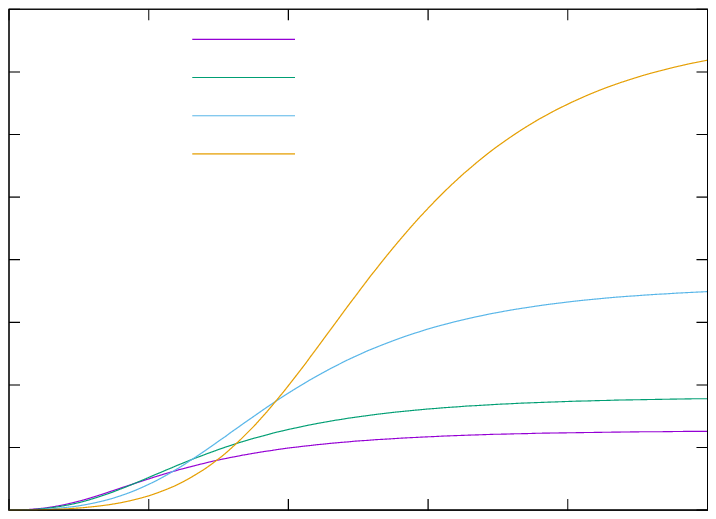}
\caption{{\it Left~:} The profile of the radial part of the electric dipole, $R_{1}(x)$, in function of $ln(1+x)$  for different values of $\zeta$ (with
increasing values of $\zeta$ from left to right) {\it Right~:} The profile of the radial part of the magnetic dipole, $S_{1}(x)$, in function of $ln(1+x)$  for different values of $\zeta$ (with increasing values of $\zeta$ from bottom to top). }
\label{fig_r1_s1}
\end{center}
\end{figure}

In the following, we will distinguish two cases:
\begin{itemize}
\item $\ell=0$: this is the monopole term. In this case we find 
\begin{equation}
R_0(x)\sim \int \frac{\sigma(x)}{x^2} \ {\rm d}x + C \ ,
\end{equation} 
where $C$ is an integration constant and the normalisation of $R_0$ can be adapted to the boundary conditions. Now using the 
behaviour of $\sigma(x)$ close to $x=0$ given by $\sigma(x\ll 1) \sim 1+ \sigma_2 x^2/2 + {\rm O}(x^3)$ (see \cite{Brihaye:2017wln}) we find~:
\begin{equation}
R_0(x\ll 1) \sim \frac{1}{x}  + C + {\rm O}(x)  \ .
\end{equation}
In order to make this more concrete, let us compute the correction to the metric up to second order in $\zeta$, which is
only valid for $\zeta << 1$. This reads~:
\begin{equation}
N=1+ x^2 + \zeta^2\left(-2 - \frac{10}{3} x^2 + \frac{2}{x} {\rm arctan}(x)\right) + {\rm O}(\zeta^{4})  \ \  , \ \
\sigma = 1 - \frac{\zeta^2}{9} \frac{1}{1+x^2} + {\rm O}(\zeta^{4})  \ .
\end{equation}
To second order in $\zeta$ Eq.~(\ref{eq1}) can be solved to give
\begin{equation}
R_0(x) \sim \left(\frac{\zeta^2}{9}-1\right) \frac{1}{x} + \frac{\zeta^2}{9} \arctan(x) \ ,
\end{equation}
where the overall normalisation can be chosen appropriately. Since the expansion only holds for small $\zeta$, the case $\zeta=3$ is excluded and the function diverges
at $x=0$ -- as confirmed by our more general numerical analysis.  

We conclude that the monopole term diverges at $x=0$  and hence {\it spherically symmetric, static and charged generalisations of the scalar-tensor solitons with approximate AdS asymptotics are not possible}. This holds true in all orders in $\zeta$ and is independent of the choice of $\zeta$.
Let us remark that there are possibilities to extend our model such that the scalar field is charged. In this case, the monopole term would be regular. We comment
on this in the conclusions.

\item $\ell > 0$~: In this case, the following boundary conditions can be employed~:
\begin{equation}
\label{r_bc}
R_{\ell}(x=0)=0  \ \ , \ \ R_{\ell}(x\rightarrow\infty)\rightarrow 1 \ \  {\rm for} \ \  \ell > 0   \ .
\end{equation}
The equation (\ref{eq1}) can only be solved numerically since the metric functions
$N$ and $\sigma$ are only known numerically. This contrasts with the case of pure AdS where analytical solutions have been constructed \cite{Herdeiro:2015vaa}. 
The modification of the profile of $R_{1}$ for different values of $\zeta$ is shown in Fig.\ref{fig_r1_s1} (left).
We find that a solution for $R_1$ exists for arbitrary values of $\zeta$. For $\zeta\rightarrow \infty$ our numerical results suggest that the derivative of the function $R_1$ at $x=0$ tends to zero for $\zeta\rightarrow \infty$. 
\end{itemize}

\subsection{Magnetic multipoles}
Choosing $A_{t}\equiv 0$, we employ the following separation of variables
\begin{equation}
A_{\varphi}(r,\theta) = S_{\ell}(r) U_{\ell}(\theta) 
\end{equation}
where for the source free exterior of the solution, the duality of the Maxwell equations implies
\begin{equation}
S_{\ell}(r)=\frac{r^2}{\sigma(r)} \frac{dR_{\ell}}{dr}  \ \ , \ \ U_{\ell}(\theta)=\sin\theta \frac{d{\cal P}_{\ell}}{d\theta}=\ell\left(\cos\theta {\cal P}_{\ell} - {\cal P}_{\ell-1}\right) \ .
\end{equation}
Note that the metric function $\sigma(r)$ enters into the relation between the magnetic and electric multipoles which
is different to the pure AdS case.
In Fig. \ref{fig_r1_s1} (right), we show the magnetic dipole function $S_1(x)$ for different values of $\zeta$. Staying with the normalization used  for $R_1(x)$ (and not, alternatively, a normalisation where 
$S_1(x\rightarrow \infty)\rightarrow 1$) we find that $S_1(x)$ tends to a constant value asymptotically and that this asymptotic value
increases with $\zeta$.

\section{Black holes}
In the case of black holes, we have used the Ansatz (\ref{ansatz}) for the scalar-tensor
field and a spherically symmetric electric field of the form
\begin{equation}
A_{\mu} dx^{\mu} = V(r) dt \ .
\end{equation}

Using this Ansatz, the equations of motion (\ref{eom1}), (\ref{eom2}) reduce to a system of four first order, ordinary differential equations
for the functions $N(r)$, $\sigma(r)$, $V'(r)$ and $\phi'(r)$, where the prime now and in the following denotes the derivative with respect to $r$. Note that the equations do not explicitly depend on $V(r)$ and $\phi(r)$ due to the shift symmetry of the system. The Maxwell equation for $V'(r)$ can be solved explicitly giving~:
\begin{equation}
\label{vsigma}
\frac{dV}{dr}\equiv V'(r)= \frac{Q \sigma(r)}{r^2}  \ ,
\end{equation}
where $Q$ is an integration constant that can be interpreted as the electric charge of the solution. The remaining equations then depend on three additional parameters~: the cosmological constant $\Lambda$, the horizon radius $r_h$ and the scalar-tensor coupling $\gamma$. By a suitable rescaling of the fields and coordinates, respectively,
we can set $\Lambda$ to a fixed value without loss of generality. 
 
In order to understand how the presence of the scalar field changes the properties of the black holes, we have
expanded the derivative of the scalar field $\phi'(r)$ in orders of $\gamma$. This reads~: 
\begin{equation}
 \phi'(r)= \gamma \phi_1'(r)+ {\cal O}(\gamma^2) 
\end{equation}
with
\begin{equation}
\phi_1'(r)= \frac{Q^4 A_4 + Q^2 A_2 + A_0}{4 r^5 r_h^2 (Q^2 + B_0)}
\end{equation}
and
\begin{equation}
A_4 = r_h (2r_h-r) \ \ , \ \ A_2 = 4r\left(\frac{\Lambda}{3}r_h^2-1\right)\left(r^3 + r_h r^2 + r r_h^2 + 3 r_h^3\right)
\end{equation}
\begin{equation}
A_0 = 16 r^2 r_h^2\left(r^2+r r_h +r_h^2\right)\left(\frac{2}{9} \Lambda^2 r^3 r_h + \frac{\Lambda^2}{9} r_h^4 - \frac{2}{3} \Lambda r_h^2+1\right) \ \ , \ \ B_0 = \frac{4}{3}\Lambda\left(r^3 r_h + r^2 r_h^2+r r_h^3\right) - 4r r_h   \ .
\end{equation}

 In the following, we will also be interested in the thermodynamic properties of the solutions, in particular in the heat capacity of the solutions for fixed values
 of $Q$ given by~:
\begin{equation}
C_Q=T_H \left(\frac{\partial S}{\partial T_H}\right)_{Q} \ ,
\end{equation}
where the entropy $S$ and the Hawking temperature $T_H$ of the solutions are given by~:
\begin{equation}
S=\frac{A_h}{4}=\pi r_h^2 \ \  \  , \  \ \  T_H=  \frac{1}{4\pi}\sigma(r_h) N'(r)\vert_{r=r_h}   \ .
\end{equation}

\subsection{Black holes for $\gamma=0$}
\label{ssection_rnads}
For $\gamma=0$ the scalar field $\phi(r)\equiv 0$ and the system of equations of motion has an analytical and well-known solution~: the Reissner-Nordstr\"om-Anti-de-Sitter (RNAdS) solution which is given by
\begin{equation}
N(r) = 1 - \frac{\Lambda}{3} r^2 - \frac{2M}{r} +  \frac{Q^2}{4r^2} \ \ , \ \  \sigma(r)\equiv 1 \ \ , \ \ 
V(r)=-\frac{Q}{r} + c  \ ,
\end{equation}
where $c$ is a constant. This solution possesses interesting thermodynamic properties when fixing the charge $Q$ of the solution \cite{Chamblin:1999tk, Chamblin:1999hg,Wu:2000id} and has been used extensively in {\it black hole chemistry} (for a review see e.g. \cite{Kubiznak:2016qmn}). We will hence remind the reader
of a number of properties of these solutions and will then use our numerical results to demonstrate how the picture changes in scalar-tensor gravity.

The mass $M$, entropy $S$ and temperature $T_H$ of the RNAdS solutions read~:
\begin{equation}
\label{RNADS}
 M=\frac{1}{2} \left(r_h + \frac{Q^2}{4r_h} - \frac{\Lambda r_h^3}{3}\right) \ \ \ , \ \ \ 
S=\frac{A_h}{4}=\pi r_h^2  \ \ \ , \ \ \  T_H=\left(\frac{\partial M}{\partial S}\right)_Q=\frac{1}{4\pi}\left(\frac{1}{r_h} - \frac{Q^2}{4r_h^3}-\Lambda r_h\right) \ \ \ , 
\end{equation}
where $r_h$ refers to the outer horizon of the black hole and $A_h$ is the area of the black hole horizon. The extremal
RNAdS with $T_H=0$ fulfils
\begin{equation}
\label{rhex}
r_{h,ex}=\sqrt{\frac{1}{2\Lambda}+\frac{1}{2\vert\Lambda\vert}\sqrt{1 - Q^2\Lambda}} \ .
\end{equation}
The heat capacity $C_Q$
at fixed charge $Q$ then reads
\begin{equation}
C_Q=\frac{32\pi^2 r_h^5 T_H}{3Q^2 - 4 r_h^2 - 4\Lambda r_h^4} \ .
\end{equation}
As pointed out in \cite{Wu:2000id}, the heat capacity $C_Q$ at fixed $Q$ 
diverges at $\partial T_H/\partial r_h = 0$ and since $\left(\partial S/\partial r_h\right) > 0$, the sign of $C_Q$ is positive (negative, respectively) depending on whether the temperature $T_H$ increases (decreases) with $r_h$. 
For $Q^2 < Q_c^2= -1/(3\Lambda)$ two values of the horizon $r_h$ exist at which $\partial T_H/\partial r_h = 0$ separating
a stable phase of small horizon black holes from a stable phase of large horizon black holes. For $Q=Q_c$ these two points merge.

\subsection{Black holes for $\gamma > 0$}
In this case, the solutions have to be constructed numerically. We have done so by using an adaptive grid collocation solver \cite{COLSYS}. In order to solve the equations numerically, we have to fix appropriate boundary conditions. 
For $r\rightarrow \infty$ we choose the following conditions~:
\begin{equation}
\sigma(r \to \infty)\rightarrow 1 \ \ , \ \ (r V)|_{r\to \infty} \rightarrow Q    \ .
\end{equation}
On the black hole horizon, we impose $N(r_h)=0$. The requirement of regularity of the scalar field on the horizon then implies
\begin{equation}
\label{eqphip_new}
f_2 (\phi'(r_h))^2 + 2 f_1 \phi'(r_h) + f_0 = 0 \ , 
\end{equation}
where $f_2$, $f_1$ and $f_0$ are expressions that contain $Q$, $r_h$, $\gamma$ and $\Lambda$ as follows:
\begin{equation}
f_2= 4\gamma r_h \Bigl(
  Q^2(2 \gamma^2 + r_h^4) + 4\Lambda   r_h^4 (r_h^4 - 2 \gamma^2) - 4 r_h^6 
\Bigr)  \ ,
\end{equation}
\begin{equation}
f_1= \Bigl( 
 \gamma^2 Q^4 + 4 \gamma^2 Q^2 r_h^2 + 2 Q^2 r_h^6 - 8 r_h^8 + 8\Lambda  r_h^4 (\gamma^2 Q^2 - 6 \gamma^2 r_h^2 + r_h^6) + 16 \Lambda^2 \gamma^2 r_h^8
 \Bigr) \ ,
\end{equation}
\begin{equation}
f_0 = \gamma r_h  
 \Bigl( -Q^4 + 24 Q^2 r_h^2 - 48 r_h^4 + 8\Lambda r_h^4 (4 r_h^2 - Q^2)
 -  16 \Lambda^2 r_h^8 \Bigr)   \ .
\end{equation}
This specific condition clearly demonstrates that solutions with {\it regular} scalar hair on the horizon do not exist for
arbitrary values of $Q$, $r_h$ and $\gamma$. We will discuss the restrictions on the parameters in \ref{subsection_restriction}.

Next to the regularity of the scalar field on the horizon, we also require the black hole temperature $T_H$ to be positive.  
For black holes with scalar-tensor hair, the Hawking temperature $T_H$ reads~:
\begin{equation}
T_H= \frac{1}{4\pi}\frac{\left(1-\frac{Q^2}{4 r_h^2} - \Lambda r_h^2\right)
\sigma(r_h)}{r_h + \gamma\phi'(r)\vert_{r=r_h}} \ ,
\end{equation}
where we have used the equations of motion.
Note that independently of the choice of $\gamma$ (finite) and as long as $\phi'(r_h)$ is finite and $\sigma(r_h)\neq 0$, the Hawking temperature becomes zero at $4\Lambda r_h^4 + Q^2 - 4 r_h^2=0$, i.e. at the extremality  condition for the RNAdS black holes. To put it differently, the Hawking temperature remains zero at the extremality condition for the RNAdS independent of $\gamma\neq 0$. Note also that $\phi'(r)\vert_{r=r_h}$ is not necessarily positive (see \ref{subsection_numerical}), while $\sigma(r_h)$ is positive. 
In general, the values of $\phi'(r)\vert_{r=r_h}$ and $\sigma(r_h)$, respectively, are only given numerically, i.e.
it is not possible -- as in the case of the RNAdS -- to give analytical expressions for $C_Q$.

\subsubsection{Restrictions on parameters}
\label{subsection_restriction}
Keeping the dependence on $\Lambda$ for the moment, the solutions to equation (\ref{eqphip_new}) read~: 
\begin{equation}
\label{phipm}
\phi'(r_h)_{\pm}  =  \frac{-f_1 \pm \sqrt{f_1^2 - f_2 f_0}}{f_2} 
                  = \frac{-f_1 \pm 2 K \sqrt{\Delta}}{f_2} 
\end{equation}
with
\begin{equation}
\label{eqK}
K \equiv |Q^2 - 4 r_h^2 + 4\Lambda r_h^4|
\end{equation}
and
\begin{equation}
 \Delta \equiv 16 \Lambda^2 \gamma^4 r_h^8 + 8 \Lambda r_h^4 \gamma^2 \left(Q^2 \gamma^2 + 4 r_h^6 - 12 \gamma^2 r_h^2\right) 
        + Q^4 \gamma^4 + 8 Q^2 \gamma^2 r_h^2\left(r_h^4 + 3 \gamma^2\right) + 4 r_h^8\left(r_h^4 - 12 \gamma^2\right)  \ .
\end{equation}
Hence, in general, we would expect two branches of the solutions (one related to $\phi'(r_h)_{+}$, the other to $\phi'(r_h)_{-}$)
to exist in this model, unless $K=0$.  Note that $K=0$ is exactly the extremality
condition for the corresponding RNAdS solutions (see (\ref{RNADS})). 

Moreover, these expressions clearly show that there exist parameters ranges for which the regularity condition (\ref{eqphip_new}) is not fulfilled.
These are~:
\begin{itemize} 
\item for $\Delta < 0$ limited by the choice of parameters that gives $\Delta=0$ for which $\phi'(r_h)_+ = \phi'(r_h)_-$,
\item for $\phi'(r_h)_{-}$ when $f_2=0$ \ .
\end{itemize}

\begin{figure}[ht!]
\input{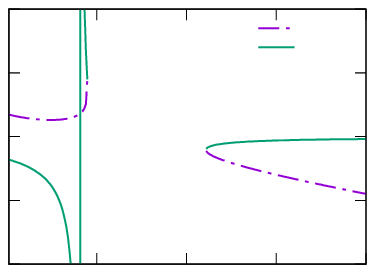}
\input{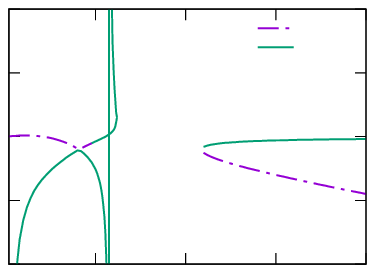}
\input{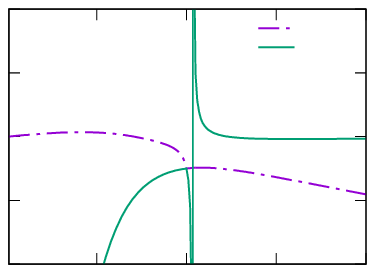}
\caption{We show the dependence of $\frac{d\phi}{dr}(r_h)\equiv \phi'(r_h)$ on the horizon value $r_h$ for $\gamma=0.2$, $\Lambda=-3$ and $Q=0$ (left), $Q=1$ (center) and $Q=4$ (right), respectively. The $+$ (purple, dashed) and $-$  (green, solid) indicate the sign appearing in (\ref{phipm}).}
\label{fig_gamma_02}
\end{figure}

To demonstrate the restriction of parameters, we have studied the value of $\phi'(r_h)$ for fixed values of $Q$ and $\gamma$ in dependence of
$r_h$. In Fig. \ref{fig_gamma_02}  we give $\phi'(r_h)$ for $\gamma=0.2$, $\Lambda=-3$ and three different values of the charge $Q$ including the uncharged case $Q=0$.
In all plots, the vertical line parallel to the $y$-axis corresponds to the parameters for which $f_2=0$. This clearly shifts to larger
values of $r_h$ with the increase of the charge $Q$. Moreover, as is clearly visible for $Q=0$ and $Q=1$, respectively, small horizon black holes
are separated by an interval of non-existence in $r_h$ from the large horizon black holes. This changes only at large values of the 
charge $Q$, here given for $Q=4$ which in the limit $\gamma=0$ corresponds to the maximal possible charge of the corresponding
RNAdS solution.

\begin{figure}[ht!]
\begin{center}
\input{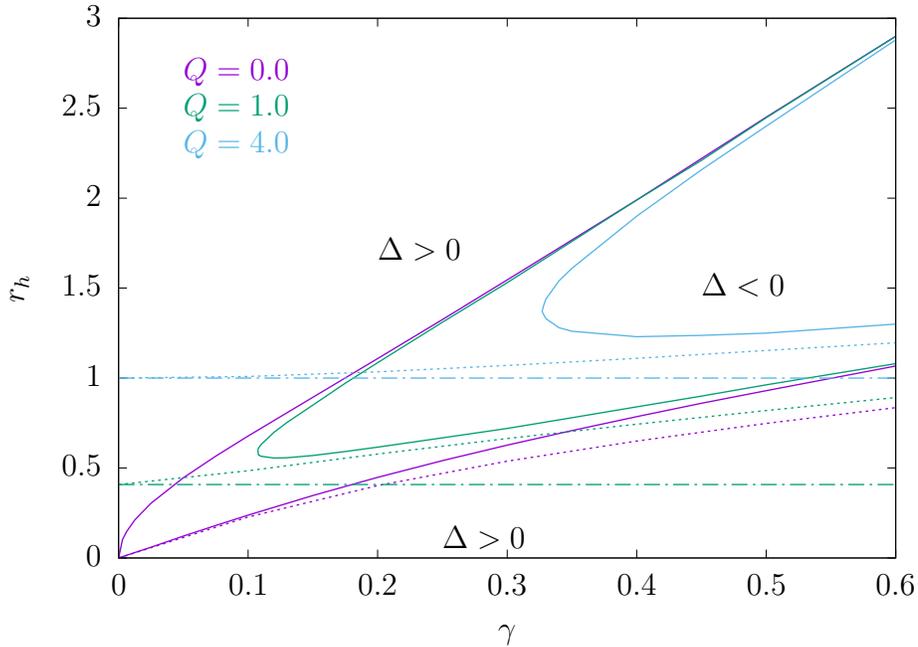}
\caption{We show the domain of existence of charged black holes in the $\gamma$-$r_h$-plane for three different values of the charge: $Q=0.0$ (purple), $Q=1.0$ (green) and $Q=4.0$ (blue). The solid lines  (with $Q$ increasing for the spike appearing at increasing values of $\gamma$)
define the lines on which $\Delta=0$ and we indicate in which part of the $\gamma$-$r_h$-plane $\Delta$ is
positive or negative, respectively. The dashed lines (with $Q$ increasing from bottom to top) correspond to $f_2=0$, while
the dotted-dashed lines (with $Q$ increasing from bottom to top) represent the values for which $K=0$ (see (\ref{phipm})). This latter condition gives the horizon
of the corresponding extremal RNAdS solution for $\gamma=0$. Black hole solutions
with scalar-tensor hair exist above the dashed line and to the left of the $\Delta=0$ lines. }
\label{fig_domain}
\end{center}
\end{figure}

As an example, we show the domain of existence of solutions with regular scalar hair for $\Lambda=-3$ and three different values of $Q$ in Fig. \ref{fig_domain}.
For $Q=0$, the condition $K=0$ does not have solutions, but for $Q=1$ and $Q=4$ the solutions of $K=0$ are $r_h=\sqrt{1/6}\approx 0.408$ and $r_h=1$,
respectively. These are the corresponding horizontal lines. The two lines that join in a spike correspond to $\Delta =0$.

\subsubsection{Numerical results}
\label{subsection_numerical}

\begin{figure}[ht!]
\begin{center}
\input{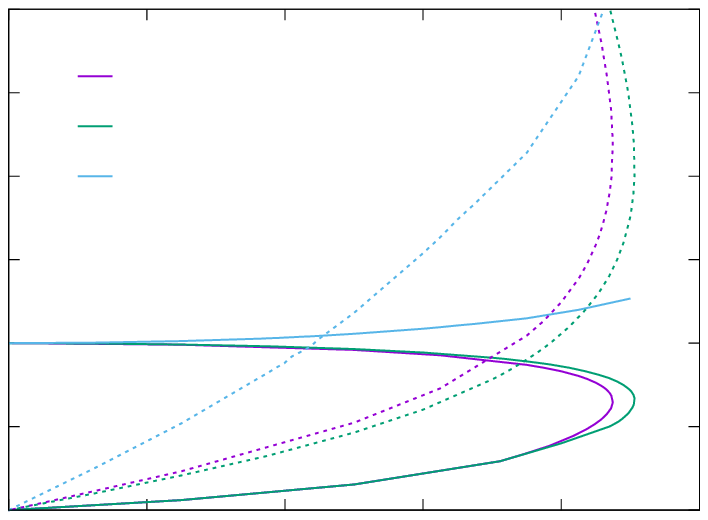}
\input{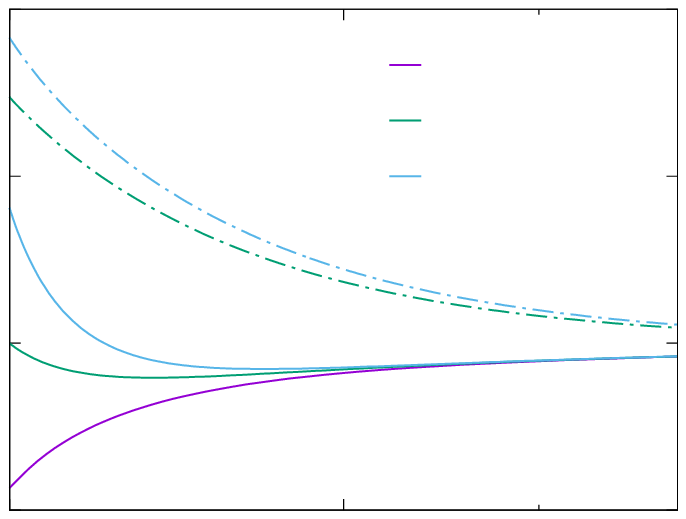}
\caption{{\it Left~:} The values of $\sigma(r_h)$ (solid) and of $\vert\frac{d\phi}{dr}(r_h)\vert$ (dashed)  for different values of the charge $Q=0.0$ (purple,  back-bending
and smaller maximal value of $\gamma$), $Q=1.0$ (green, back-bending, larger maximal value of $\gamma$), $Q=3.5$ (blue, no back-bending) in dependence of the scalar-tensor coupling $\gamma$ for $r_h=1.0$ and $\Lambda=-3$.  {\it Right~:} The profiles of the derivative of the scalar field function $\frac{d\phi}{dr}$ (solid with $Q$ increasing from bottom to top) and of the derivative of the gauge field function $\frac{dV}{dr}$ (dotted-dashed with $Q$
increasing from bottom to top) for $r_h=1.0$, $\gamma=0.1$, $\Lambda=-3$ and three values of the charge $Q=0.0$ (purple), $Q=2.5$ (green) ad $Q=3.2$ (blue), respectively.}
\label{fig_BH_data}
\end{center}
\end{figure}

 \paragraph{Uncharged black holes}
 Here, we briefly review the uncharged case $Q=0$. This has been studied previously  \cite{Brihaye:2017wln}, however,
 we present new results here that we believe to be important in order to understand the charged case. $Q=0$ implies 
 $V(r)\equiv 0$. 
As was shown in \cite{Brihaye:2017wln}, the scalar-tensor black holes exist up to a maximal value of the scalar-tensor coupling
$\gamma=\gamma_{\rm max}$ (see Fig. \ref{fig_domain} for $Q=0$). 
This branch connects to the Schwarzschild-AdS (SAdS) solution in the limit $\gamma=0$ (with value $\sigma(r_h)=1$) and can be constructed by continuous 
deformation of the latter. We will refer to this as the {\it main branch} in the following. 
In Fig.~\ref{fig_BH_data} (left) we show the behaviour of some parameters characterising the solution with $r_h=1.0$ and $\Lambda=-3$, i.e. 
the values of the metric function $\sigma(r)$ at the horizon, $\sigma(r_h)$, and the value of the derivative
of the scalar field function $\phi(r)$ at the horizon, $\vert\phi'(r_h)\vert$, as function of $\gamma$. 
In Fig.~\ref{fig_BH_data} (right) we show the profile of $\phi'(r)$ for $r_h=1.0$ and $\gamma=0.1$. 
As is obvious from this figure, $\phi'(r_h) < 0$. 

Apart from this first branch of solutions we have been able to construct a second branch of solutions that starts at $\gamma_{\rm max}$ and extends back in
$\gamma$ reaching a singular limit for $\gamma\rightarrow 0$. This singular solution has $\phi'(r_h)\rightarrow \infty$ and
$\sigma(r_h)\rightarrow 0$, see Fig.~\ref{fig_BH_data} (left).
For a fixed value of $\gamma$, the solutions on the second branch have larger values of $\phi'(r_h)$ and smaller values of $\sigma(r_h)$ than the corresponding solution on the first branch. 

As shown in Fig.~\ref{fig_domain} the value  $\gamma_{max}$ decreases strongly when the horizon radius $r_h$ is decreased. 
We find $\gamma_{\rm max}\approx 0.1748$ for $r_h=0.1$,  while for $r_h=0.5$ and $r_h=1.0$ our numerical
analysis gives $\gamma_{\rm max}\approx 0.0610$ and $\gamma_{\rm max}\approx 0.0029$, respectively. 
To state it differently~:  when the chosen horizon radius $r_h$ tends to the value of the corresponding AdS radius
of a space-time without the scalar field $r^2_{\rm AdS}=-3/\Lambda$, which with our choice of $\Lambda=-3$ throughout the
paper equates to $r^2_{\rm AdS}=1$, the interval in $\gamma$ for which scalar-tensor black holes exist shrinks. 
Note also that the presence of the scalar field allows to construct black holes with $r_h=r_{\rm AdS}$.

\paragraph{Charged black holes}
The main difference to the uncharged case can be seen in Fig.~\ref{fig_domain},  where we give the domain
of existence of solutions in the $r_h$-$\gamma$-plane for two different values of non-vanishing $Q$. As can be clearly seen,
the $\Delta=0$ line, which restricts the existence of the solutions, does not extend to small horizon values anymore. 
In order to explain the pattern, also with view to \ref{subsection_restriction}, let us discuss cases with fixed value of $Q$.
As mentioned above, the extremality condition of the RNAdS black holes equal that for $K=0$. In the following, we will
refer to the horizon value that fulfils $K=0$ as to $r_{h,0}\equiv r_{h,ex}$ (see (\ref{rhex})). 
The pattern can then be described as follows~:

\begin{itemize}
\item for $r_h \gg r_{h,0}$ the RNAdS solution gets progressively deformed when increasing $\gamma$ from zero, forming a branch
of black holes with scalar-tensor hair limited by the  maximal value of $\gamma=\gamma_{\rm max}$ which corresponds to $\Delta = 0$. An example in Fig.~\ref{fig_domain} would be $Q=1.0$, $r_h=1.0 \gg r_{h,0}\approx 0.408$. The corresponding data for this
case is shown in Fig.~\ref{fig_BH_data} (left). Again, we find two branches of solutions and a back-bending behaviour for
$\phi'(r_h)$, very similar to the case with $Q=0$. 
  
\item for $r_h \gtrsim r_{h,0}$ the branch of scalar-tensor black holes can be constructed by deformation
of the RNAdS solutions as above, however, before reaching the $\Delta=0$ curve, the branch reaches the
line $f_2=0$ at some $\gamma=\gamma_{cr}$. A typical case would be $r_h=1.0$ and $Q=4.0$ in Fig.~\ref{fig_domain}, for which $r_{h,0}=r_{h,ex}=1.0$. In order to understand what happens when increasing $Q$ 
and with that the value of $r_{h,0}$, we show $\vert\phi'(r_h)\vert$ as well as $\sigma(r_h)$  in 
Fig.~\ref{fig_BH_data} (left) for $Q=3.5$.  The second branch is absent in this case and no backbending exists.
From Fig.~\ref{fig_BH_data} (right) it also becomes clear what happens when increasing $Q$. 
When increasing $Q$, we find that $\phi'(r_h)$ becomes positive at some intermediate value of the charge, 
here $Q=2.5$ and increases with the further increase of $Q$, see the profile for $Q=3.2$. On the other hand, $V'(r_h)$ is always positive. The profiles further suggest that when increasing the charge $Q$ the derivative of $\phi'(r)$ close to the horizon increases strongly. 

\item for $r_h <  r_{h,0}\equiv r_{h,ex}$ we have not been able to construct solutions, i.e. scalar-tensor black holes
(seem to) exist only for horizon value larger than the corresponding horizon value of the extremal RNAdS solution.

\end{itemize}

\begin{figure}[ht!]
\begin{center}
\input{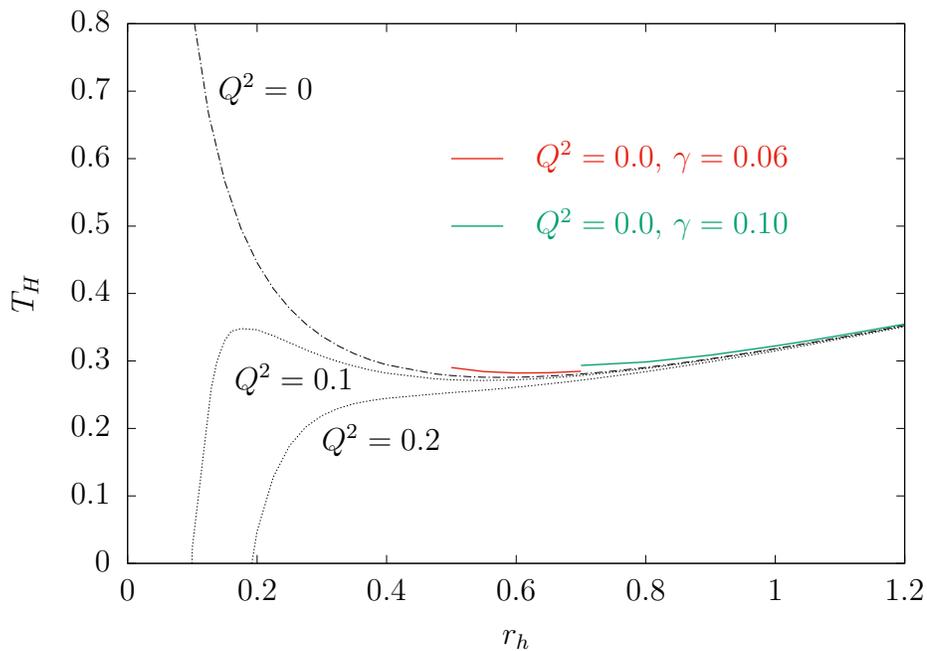}
\caption{We show the Hawking temperature $T_H$ in dependence of $r_h$ for
a Scharzschild-Anti-de Sitter black hole ($Q^2=0$, dotted-dashed black) and two differently charged
RNAdS black holes (solid black) with charges $Q^2=0.1$ and $Q^2=0.2$, respectively. We also give $T_H$ for two uncharged black holes with scalar-tensor hair for $\gamma=0.06$ (solid red,  smaller rminimal value of $r_h$) and $\gamma=0.1$ (solid green, larger minimal value of $r_h$). }
\label{fig_TH}
\end{center}
\end{figure} 

With the domain of existence of solutions at hand, we have then studied the heat capacity $C_Q$ of these solutions.
Our results are shown in Fig. \ref{fig_TH} for $Q^2=0$ and two different values of $\gamma$. The curves
for $Q^2\neq 0$ are indistinguishable from the ones shown, that is why we do not present them here.
The first obvious difference to the case of the uncharged RNAdS case, which corresponds to a Schwarzschild-AdS (SAdS) solution, as well as the RNAdS case (see curves for $Q^2=0.1$ and $Q^2=0.2$) is that scalar-tensor black holes exist only on a finite interval of the 
horizon radius $r_h$. In particular, as outlined above, small horizon black holes do not exist. 
In particular, we find that for $\gamma$ large enough, $C_Q$ remains positive on the full branch of solutions.
This is indicated by the $Q^2=0$, $\gamma=0.1$ curve. On the other hand, for $\gamma$ small we observe that
a transition from $C_Q < 0$ to $C_Q > 0$ with $C_Q\rightarrow \infty$ (or equivalently $\partial T_H/\partial r_h=0$) at $r_{h,crit}$ exists. This is indicated by
the curve for $Q^2=0$ and $\gamma=0.06$. While $r_{h,crit}=\sqrt{-1/\Lambda}$ for the SAdS case, which is $r_{h,crit}=\sqrt{1/3}\approx 0.58$ for our choice of $\Lambda=-3$, we find that for $\gamma=0.06$, the
value of $r_{h, rit}\approx 0.60$. This means that the critical value of the horizon radius $r_h$ at which a 
transition from thermodynamically unstable ($C_Q < 0$) to thermodynamically stable ($C_Q > 0$) black holes
appears shifts to larger values of $r_h$.

\section{Conclusions and Outlook}
Whenever alternative theories of gravity are studied, black holes that possess non-trivial
matter fields, e.g. scalar fields, on the horizon appear. Hence, for these black holes there is no equivalent of the No-hair theorems that exist for
electro-vacuum black holes in 4 dimensional asymptotically flat space-time within General Relativity (GR).
In this paper, we have studied a scalar-tensor extension of GR that
contains a higher order curvature correction, the Gauss-Bonnet term,  non-minimally coupled to a real scalar field.
The resulting coupled, non-linear field equations of this model can only be solved numerically.
The family of solutions obtained is a generalisation of the RNAdS black holes in the sense that
the solutions are not only characterised by their charge, cosmological constant and horizon radius (or equivalently their mass parameter $M$), but by an additional parameter, the scalar-tensor coupling constant $\gamma$.
We find that the domain of existence of these solutions shows a very complicated pattern and that 
solutions exist only in a finite domain of the $r_h$-$Q$-$\gamma$ parameter space for fixed value of the
cosmological constant. We find parameters choice for which zero, one or two solutions exist.
Critical values of the parameters beyond which no scalar-tensor black holes exist  can be characterised by analytic expression
and are related  to the regularity of the scalar field at the black hole horizon. 
In particular, we find that small black holes, i.e. solutions with small horizon values, do not exist in our model.

While we have fixed the cosmological constant within this work, another approach would be to vary the cosmological constant. Then, the existence of
phases of black holes in dependence of the pressure $p=-\Lambda/(8\pi)$ could be studied
and compared to phase transitions in chemical systems \cite{Kubiznak:2014zwa}. Moreover, the Einstein-Maxwell-AdS solutions
themselves can be extended to include arbitrary electric multipoles as shown in \cite{Herdeiro:2016plq}. It would be interesting
to see whether the corresponding black holes with scalar hair can exist. This is currently under investigation.

Let us finally state that our model can be extended to a complex scalar field charged under a local U(1). For that, however, the scalar-tensor term would have to be changed to $\vert\phi\vert {\cal G}$  in order to guarantee both a real Lagrangian density and the invariance under $U(1)$ phase transformation.
This would, however, contain a square root of the two scalar degrees of
freedom coupled to the Gauss-Bonnet term which does not seem very natural. To avoid this problem, one could replace the linear term
by a quadratic one, i.e. a term of the form $\vert\phi\vert^2 {\cal G}$. This model has very recently been studied in \cite{Brihaye:2018grv}
and it has been demonstrated that spherically symmetric boson stars exist. These are globally regular, which demonstrates that
the divergence of the monopole term for the case of an uncharged scalar field can be avoided when charging the scalar field
and replacing the linear scalar-tensor coupling term by a quadratic coupling term.

\vspace{1cm}

{\bf Acknowledgements} 
 BH would like to thank FAPESP for financial support under
grant number {\it 2016/12605-2} and CNPq for financial support under
{\it Bolsa de Produtividade Grant 304100/2015-3}.  

\clearpage



 \end{document}